\documentclass[aps,twocolumn,amsmath,amssymb,showpacs,prl,superscriptaddress,unsortedaddress]{revtex4}

\usepackage{epsf}
\usepackage{graphicx}
\usepackage{sidecap}
\usepackage{amsmath,amsfonts,amssymb}

\begin{document}

\title{Anomalous spin-momentum locked two-dimensional states in the vicinity of a topological phase transition}

\author{Su-Yang~Xu}
\affiliation {Joseph Henry Laboratory and Department of Physics, Princeton University, Princeton, New Jersey 08544, USA}

\author{M.~Neupane}
\affiliation {Joseph Henry Laboratory and Department of Physics, Princeton University, Princeton, New Jersey 08544, USA}

\author{Chang~Liu}
\affiliation {Joseph Henry Laboratory and Department of Physics, Princeton University, Princeton, New Jersey 08544, USA}

\author{S.~Jia}
\affiliation {Department of Chemistry, Princeton University, Princeton, New Jersey 08544, USA}

\author{L.~A.~Wray}
\affiliation{Joseph Henry Laboratory and Department of Physics, Princeton University, Princeton, New Jersey 08544, USA}
\affiliation{Advanced Light Source, Lawrence Berkeley National Laboratory, Berkeley, California 94305, USA}

\author{G.~Landolt}
\affiliation {Swiss Light Source, Paul Scherrer Institute, CH-5232, Villigen, Switzerland}
\affiliation {Physik-Institute, Universitat Zurich-Irchel, CH-8057 Zurich, Switzerland}

\author{B.~Slomski}
\affiliation {Swiss Light Source, Paul Scherrer Institute, CH-5232, Villigen, Switzerland}
\affiliation {Physik-Institute, Universitat Zurich-Irchel, CH-8057 Zurich, Switzerland}

\author{J.~H.~Dil}
\affiliation {Swiss Light Source, Paul Scherrer Institute, CH-5232, Villigen, Switzerland}
\affiliation {Physik-Institute, Universitat Zurich-Irchel, CH-8057 Zurich, Switzerland}

\author{N.~Alidoust}
\affiliation {Joseph Henry Laboratory and Department of Physics, Princeton University, Princeton, New Jersey 08544, USA}

\author{S.~Basak}
\affiliation {Department of Physics, Northeastern University, Boston, Massachusetts 02115, USA}

\author{H.~Lin}
\affiliation {Department of Physics, Northeastern University, Boston, Massachusetts 02115, USA}

\author{J.~Osterwalder}
\affiliation {Physik-Institute, Universitat Zurich-Irchel, CH-8057 Zurich, Switzerland}

\author{A.~Bansil}
\affiliation {Department of Physics, Northeastern University, Boston, Massachusetts 02115, USA}

\author{R.~J.~Cava}
\affiliation {Department of Chemistry, Princeton University, Princeton, New Jersey 08544, USA}

\author{M.~Z.~Hasan}
\affiliation {Joseph Henry Laboratory and Department of Physics, Princeton University, Princeton, New Jersey 08544, USA}

\date{\today}

\begin{abstract}
We perform spin-resolved and spin-integrated angle-resolved photoemission spectroscopy measurements on a series of compositions in the BiTl(S$_{1-\delta}$Se$_\delta$)$_2$ system, focusing on $\delta$-values in the vicinity of the critical point for the topological phase transition (the band inversion composition). We observe quasi two dimensional (2D) states on the outer boundary of the bulk electronic bands in the trivial side (non-inverted regime) of the transition. Systematic spin-sensitive measurements reveal that the observed 2D states are spin-momentum locked, whose spin texture resembles the helical spin texture on the surface of a topological insulator. These anomalous states are observed to be only prominent near the critical point, thus are possibly related to strong precursor states of topological phase transition near the relaxed surface.
\end{abstract}


\maketitle

Three dimensional topological insulator is a novel phase of matter that has attracted much interests in the condensed matter physics community \cite{Moore Nature insight, Fu Liang PRB topological invariants, David Science BiSb}. It is demonstrated that a topological phase transition (TPT) from a trivial insulator to a topological insulator occurs by replacing sulfur with selenium in the BiTl(S$_{1-\delta}$Se$_\delta$)$_2$ system \cite{Suyang, Sato}. Using photoemission probes, previous works \cite{Suyang, Sato} measured the electronic structure of the BiTl(S$_{1-\delta}$Se$_\delta$)$_2$ system with relatively large $\delta$ step ($\delta$ step $\sim0.2$), from which they observed that the band gap closes and reopen again (the so-called band inversion process) upon increasing $\delta$ and therefore built up the frame of the TPT in the BiTl(S$_{1-\delta}$Se$_\delta$)$_2$ system.

Although the frame has been built, the systematic and detailed properties of the band inversion process and the TPT in the BiTl(S$_{1-\delta}$Se$_\delta$)$_2$ system remain completely unknown both theoretically and experimentally. For example, one aspect that is of particular interest is the precise chemical composition of the quantum critical point ($\delta_c$ where bulk band gap becomes zero) of the phase transition. The system at $\delta_c$ has been theoretically understood to have a 3+1 dimensional massless Dirac fermion \cite{Pallab}, which is predicted to be multiply connected to many other undiscovered novel topological phases such as a topological Weyl semimetal \cite{Gil, Weyl Balents}. In fact, it is even experimentally unknown whether the TPT in the BiTl(S$_{1-\delta}$Se$_\delta$)$_2$ system is a sharp transition as expected in the simplest theoretical picture \cite{Fu Liang PRB topological invariants}, or the transition is intrinsically broadened so that no precise critical point $\delta_c$ can be determined. An ideal physical property, in order to investigate such physical process, is the \textit{spin} structure across the TPT. In the sharp transition case, the spin structure is expected to show a sudden change from doubly spin degenerate (topologically trivial regime) to spin polarized (topological insulator regime). However, if the transition in BiTl(S$_{1-\delta}$Se$_\delta$)$_2$ is indeed broadened, then the change of the spin channel across the TPT is completely unknown. In fact, it is possible that the spin transition is also softened in certain ways. For example the electronic structure even on the topologically trivial side can pick up spin polarization prior to the phase transition.

\begin{SCfigure*}
\centering
\includegraphics[width=13.5cm]{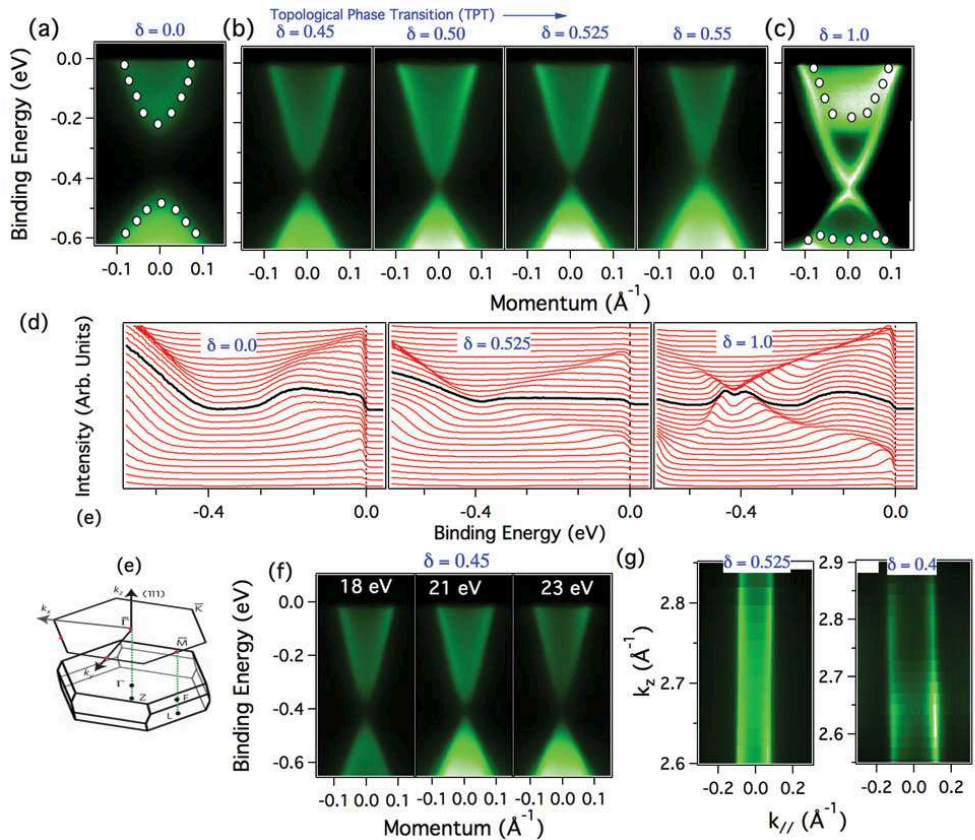}
\caption{Observation of quasi two dimensional states on the trivial side of the topological phase transition. (a)-(c) ARPES $k$-$E$ maps of BiTl(S$_{1-\delta}$Se$_\delta$)$_2$ for (a) trivial insulator, $\delta$ = 0.0, (b) concentrations near the band inversion, $\delta$ = 0.45, 0.50, 0.525 and 0.55, and (c) topological insulator, $\delta$ = 1.0. ARPES spectra are measured at the $\bar{\Gamma}$ point along the $\bar{\Gamma}$-$\bar{K}$ direction. White dots are guides to the eye for locations of the conduction and valence bands. Spectrum for $\delta$ = 0.45 is highlighted. (d) Energy distribution curves (EDCs) for $\delta$ = 0.0, 0.525, and 1.0. The EDC through $\bar{\Gamma}$ point is highlighted. (e) A schematic diagram of the three dimensional Brillouin zone of BiTl(S$_{1-\delta}$Se$_\delta$)$_2$ and the two dimensional Brillouin zone for the projected (111) surface. (f) Photon energy dependence spectra for $\delta$ = 0.45. (g) $k_z$ dispersion maps for $\delta$ = 0.525 and 0.4.}
\end{SCfigure*}

In this Letter, we make the experimental efforts to determine these interesting physical properties along the TPT. Using angle-resolved photoemission spectroscopies (ARPES) and spin-resolved ARPES, we present systematic electronic and spin structure measurements on BiTl(S$_{1-\delta}$Se$_\delta$)$_2$ with $\delta=0.0, 0.4, 0.45, 0.5, 0.525, 0.55, 1.0$ respectively. Single crystals of BiTl(S$_{1-\delta}$Se$_\delta$)$_2$ with various $\delta$ used for this measurements were grown using the Bridgman method, which is described elsewhere \cite{Jia}. High resolution (10-25 meV) ARPES measurements of the low-energy electronic structure were performed at the Synchrotron Radiation Center (SRC), Stanford Synchrotron Radiation LightSource (SSRL) and Advanced Light Source (ALS). Spin-resolved ARPES \cite{Hugo Review} measurements were performed at the SIS beamline of the Swiss Light Source, Switzerland using the COPHEE spectrometer with two 40 kV classical Mott detectors. Samples were cleaved {\it in situ} and measured at 10-80 K in a vacuum better than 1 $\times$ 10$^{-10}$ torr. To calculate the $k_z$ of the photon energy dependence measurements, we use the relation $k_z=\frac{1}{\hbar}\sqrt{2m(E_{kin}\cos^2\theta+V_0)}$ with inner potential $V_0$$\sim$12 eV. The theoretical band calculations were performed with the LAPW method using WIEN2K package \cite{Blaha} within the framework of the density functional theory (DFT). The generalized gradient approximation (GGA) was used to model exchange-correlation effects. For semi-infinite surface system, calculations are done based on a Green's function with implementation of the experimentally-based $k \cdot p$ model in Refs. \cite{LiFu1, LiFu2} to reveal the surface spectral weight and spin texture of BiTl(S$_{1-\delta}$Se$_\delta$)$_2$.

We start by showing electronic structure (spin-integrated) of the BiTl(S$_{1-\delta}$Se$_\delta$)$_2$ system at different $\delta$. As shown in Fig. 1(a) and (c), the two end compounds ($\delta=0.0$ and $1.0$) clearly show the absence and the presence of Dirac surface states connecting the bulk conduction and valence bands, proving the trivial insulator ($\delta=0.0$) and the topological insulator ($\delta=1.0$) phase respectively \cite{Suyang}. The trivial insulator state has been shown to extend from $\delta=0.0$ to $0.4$, whereas the topological state is observed from $\delta=1.0$ to $0.6$ (see Ref. \cite{Suyang} and the supplementary information). Now we focus on fine tuning of compositions in between $0.4$ and $0.6$, as shown in Fig. 1(b). A small (but still clearly observable) band gap of about 30 meV is observed on $\delta=0.45$ in Fig. 1(b), indicating that the system still belongs to the trivial insulator regime. When increasing $\delta$ to $0.50$ and $0.525$, the bulk conduction and valence bands are observed to further approach each other, and the gap is not observable in the dispersion maps [Fig. 1(b)]. The overall measured electronic structure at $\delta=0.50$ or $0.525$ in Fig. 1(b) shows linear (Dirac-like) dispersion. At $\delta=0.60$ (shown in supplementary), we observe that the bulk conduction and valence bands are separated again by an observable band gap with the surface states connecting the band gap, demonstrating that the system belongs to the topological insulator regime. The topological nature of the system at $\delta=0.50$ or $0.525$ (whether trivial or nontrivial) can not be unambiguously determined by the dispersion measurements in Fig. 1(b). The linear (Dirac-like) band can be the two dimensional topological surface states or the three dimensional bulk Dirac states \cite{Pallab}.

\begin{figure}
\centering
\includegraphics[width=8.5cm]{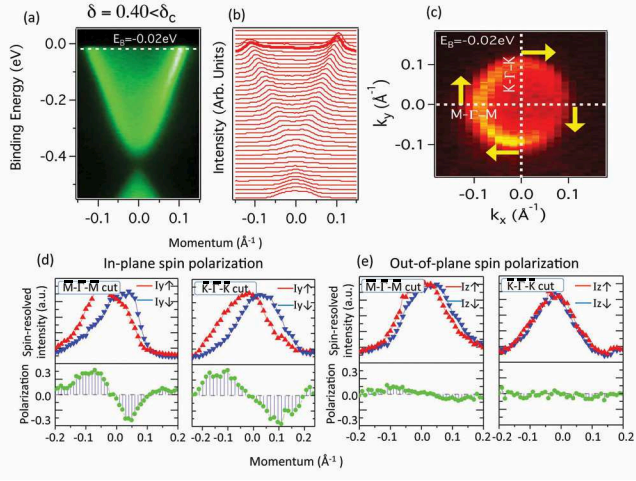}
\caption{Observation of spin-momentum locked states on the trivial side of the topological phase transition. (a) ARPES $k$-$E$ map of BiTl(S$_{1-\delta}$Se$_\delta$)$_2$ for $\delta=0.40$. Dotted line shows the binding energy where the spin-resolved measurements are performed [(d)-(e)]. (b) Momentum distribution curves (MDCs) of the spectrum in (a). Highlighted MDC curve is chosen for spin-resolved measurement. (c) Fermi surface mapping for $\delta=0.40$. The two spin-resolved momentum distribution curves (SR-MDC) measurements are along $\bar{\Gamma}$-$\bar{M}$ and $\bar{\Gamma}$-$\bar{K}$ directions respectively. Yellow arrows represent the measured spin polarization vector around the Fermi surface. (d) Upper panel (left/right): measured SR-MDC along $\bar{\Gamma}$-$\bar{M}$/$\bar{\Gamma}$-$\bar{K}$ direction.  Lower panel (left/right): measured in plane spin polarization along $\bar{\Gamma}$-$\bar{M}$/$\bar{\Gamma}$-$\bar{K}$ direction. (e) same as (d) but for out-of-plane spin component.}
\end{figure}

Now we measure the electronic dispersion as a function of incident photon energy. By varying the photon energy, one can effectively probe the electronic structure at different out-of-plane momentum $k_z$ values in a three dimensional Brillouin zone [Fig. 1(e)]. Fig. 1(g) shows the photon energy ($k_z$) measurements at $\delta=0.525$ by Fermi surface mapping in $k_\|$ vs $k_z$ momentum space. The straight Fermi lines in parallel with $k_z$ direction show the absence of observable $k_z$ dispersion (for systematic measurements, see supplementary information), which seems to suggest that the Dirac bands at $\delta=0.525$ are topological surface states. However, far from the end of the story, we have also carried out photon energy dependence measurements at $\delta=0.45$, where there is a clear band gap. As shown in Fig. 1(f) and (g), even for $\delta=0.45$, no clear $k_z$ dispersion is observed, which indicates the quasi two dimensional nature of the observed bands even in the trivial insulator regime ($\delta=0.45$ or 0.40). Thus at $\delta=0.45$ or $0.40$, we observe a two dimensional state that marks the outer boundary of the bulk bands. The ``true'' bulk bands deep inside the bulk crystals is, however, not clearly resolved in our measurements. This is because ARPES is a surface-sensitive probe (penetration depth within $10$ $\mathrm{\AA}$, see supplementary information). In narrow band gap semiconductor systems, the surface is typically subject to surface relaxation and band bending effects \cite{David Nature tunable, vdW}, which trap the bulk electronic states near the surface, and cause additional two dimensional electronic states \cite{Andrew Nature physics Fe, Hofmann}. These effects make the ``true'' bulk bands deep inside the bulk crystal not resolved by ARPES. The fact that the observed electronic states at $\delta=0.45$ or $0.40$ are dominated by the two dimensional states at the outer boundary of the bulk bands must hinder us from precise pinpointing the composition of the critical point $\delta_c$. Such observation implies that on the relaxed surface of BiTl(S$_{1-\delta}$Se$_\delta$)$_2$ system, the evolution of the electronic structure across the TPT is far different than the simplest theoretical picture where the electronic structure is expected to go from 3+1 dimensional massive Dirac bands (non-inverted trivial regime, fully gapped), to 3+1 dimensional massless Dirac bands (critical point, gapless), and then to 3+1 dimensional massive Dirac bands (inverted nontrivial regime, topological surface states connecting the bulk gap) \cite{Pallab}.

\begin{figure}
\centering
\includegraphics[width=8.5cm]{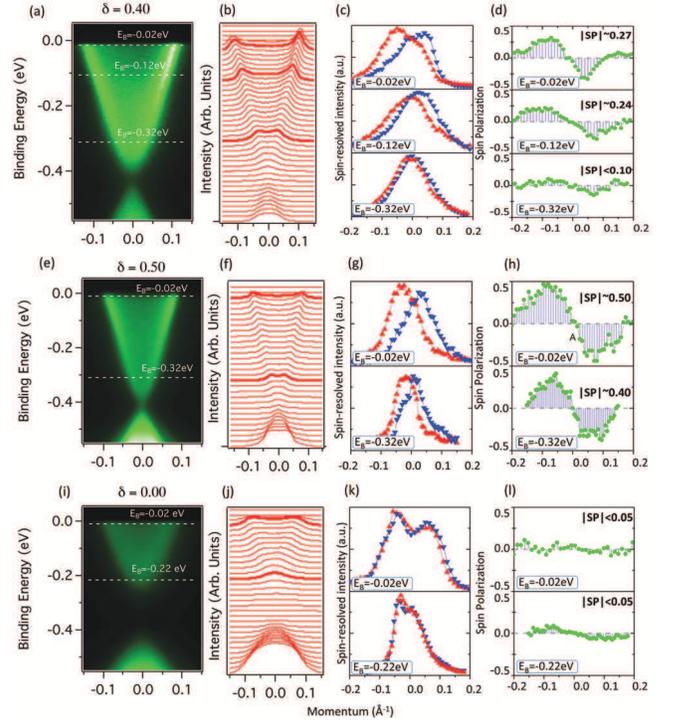}
\caption{Evolution of spin-momentum locked states with binding energy and doping. (a),(e),(i) ARPES $k$-$E$ maps with dotted line indicating the energy levels of spin-resolved MDC measurements. Doping levels are marked at the top of each map. (b),(f),(j) MDCs with highlighted curves chosen for spin-resolved measurement. (c),(g),(k) spin-resolved MDC spectra. (d),(h),(l) corresponding spin polarization measurements.}
\end{figure}

\begin{figure*}
\centering
\includegraphics[width=15cm]{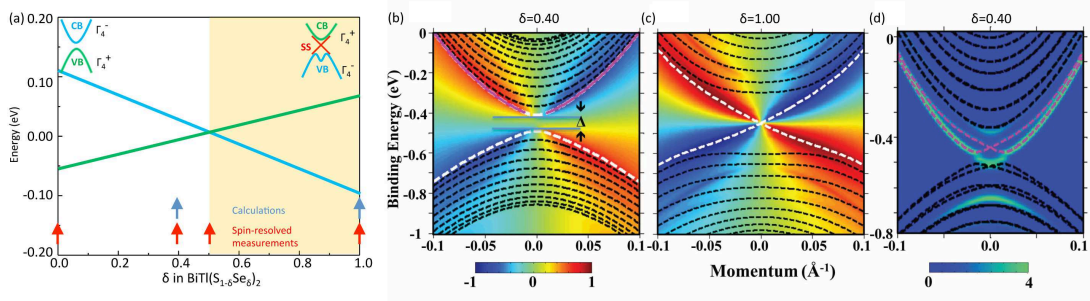}
\caption{Theoretical calculation of spin-resolved spectral weight on the surface of BiTl(S$_{1-\delta}$Se$_\delta$)$_2$. (a) The energy levels of the bulk conduction and valence bands on the two ends ($\delta=0.0$ and $1.0$) are connected by straight lines to show the general trend of the band inversion. The compositions chosen for spin-resolved measurements in Fig. 2 and 3 and calculations in Fig. 4 are marked by red and blue arrows respectively. (b) Theoretical simulations of the bulk and surface dispersions on a semi-infinite system of BiTl(S$_{1-\delta}$Se$_\delta$)$_2$ in the trivial phase prior to the quantum phase transition. (c) Same as (b) but for the topological insulator phase. Colors represent the spin polarization calculated by taking the difference between the up and down spin and dividing them by their sum [$(I_{\mathrm{up}}-I_{\mathrm{down}})/(I_{\mathrm{up}}+I_{\mathrm{down}})$]. Here `up' spin is along the positive $k_y$-direction and vice versa. (c) Energy bands in presence of an external potential at the surface. Color displays their spectral weight. Magenta lines represent the surface Rashba cones now shifted inside the energy gap under the influence of the external potential.}
\end{figure*}

Now we perform spin-resolved measurements on the newly observed two dimensional states on the trivial side, in order to test if such states are spin polarized or not. We start from $\delta=0.40$ [Fig. 2(a)] and focus on the vicinity of the Fermi level ($E_B=-0.02eV$). The momentum distribution curves (MDCs) for the spectrum in Fig. 2(a) are shown in Fig. 2(b), where the highlighted curve is chosen for spin-resolved MDC measurement. Left panels of Fig. 2(d) show the in-plane spin-resolved momentum distribution curve (SR-MDC) spectra as well as the measured in-plane spin polarization along $\bar{\Gamma}$-$\bar{M}$ direction. Clear in-plane spin polarization is observed on the two dimensional states from Fig. 2(d). Furthermore, the measured spin polarization in Fig. 2(d) (left) shows that the spin texture is arranged in the way that spins have opposite directions on the opposite sides of the Fermi surface. In order to map out the in-plane spin texture around the Fermi surface [Fig. 2(c)], SR-MDC measurement is performed along another momentum direction $\bar{\Gamma}$-$\bar{K}$. As shown in Fig. 2(e) right panels, the spin-resolved measurement reveal the same spin-momentum locking configuration along $\bar{\Gamma}$-$\bar{K}$ direction. Corresponding out-of-plane spin polarization along $\bar{\Gamma}$-$\bar{M}$ and $\bar{\Gamma}$-$\bar{K}$ is shown in Fig. 2(e). No significant out-of-plane spin polarization [Fig. 2(e)] is observed within our experimental resolution. The spin texture configuration obtained from these spin-resolved measurements in Fig. 2(d) and (e) is schematically shown by the arrows in Fig. 2(c). Surprisingly, our spin-resolved measurements reveal that these two dimensional states are not only spin polarized, but also their spin texture resembles the helical spin texture on the topological surface states as observed in Bi$_2$Se$_3$ \cite{David Nature tunable} and BiTl(S$_0$Se$_1$)$_2$ ($\delta=1.0$) \cite{Suyang} from the Fermi surface point of view.

After showing the spin configuration of these two dimensional states at $\delta=0.40$, we present systematic spin-resolved studies to show how the spin texture evolves as a function of binding energy $E_B$ and composition $\delta$. We begin with $\delta=0.40$ and show spin-resolved measurements at different binding energies. As shown in Fig. 3(a)-(d), the spin-momentum locking behavior is observed at all binding energies from near the Fermi level $E_B=-0.02$ eV to energy near the conduction band minimum $E_B=-0.32$ eV. While the magnitude of the spin polarization on the Fermi level is found to be around 0.3, the spin polarization magnitude is found to decrease to almost zero when approaching small momenta near the Kramers' point $\bar{\Gamma}$ (the conduction band minimum). Now we turn to the gapless case with the system at $\delta=0.50$. Again, spin-resolved measurements [Fig. 3(e)-(h)] reveal the same spin texture configuration, which resembles the helical spin texture of the topological surface states on the topologically nontrivial side. The magnitude of the spin polarization is around 0.5 on the Fermi energy, and does not show obvious reduction of magnitude when going to small momentum near the Kramers' point $\bar{\Gamma}$ (spin polarization $\sim$ 0.45 at the Dirac point, $E_B=-0.32$ eV). Finally, we turn to the composition far in the topologically trivial side ($\delta=0.0$). Our spin-resolved measurements [Fig. 3(i)-(l)] show only very weak polarizations ($\sim 0.05$). The magnitude of the spin polarization is too weak (comparable to the instrumental resolution) to figure out the spin texture configuration around the Fermi surface at $\delta=0.0$ (We noticed an independent work also reports the near absence of spin polarization at $\delta=0.0$ in the BiTl(S$_{1-\delta}$Se$_\delta$)$_2$ system \cite{Ando}.)

In order to cross-check the results obtained from spin-resolved measurements, we have also created a Green's function implementation of the experimentally-based $k \cdot p$ model in Refs. \cite{LiFu1,LiFu2} to simulate the bulk and surface dispersions of a semi-infinite crystal of BiTl(S$_{1-\delta}$Se$_\delta$)$_2$. The spin-resolved surface dispersions are shown in the colorplots of Fig. 4(b) and (c) for trivial and non-trivial phases, respectively, the maximum intensity being the up spin and the minimum being the down spin along the $y$-direction. The black dashed lines are the bulk bands obtained from a slab calculation. The white dashed lines represent the edge of the bulk bands in Fig. 4(b) and the surface states in Fig. 4(c). To determine the nature of the electronic states near the surface on the trivial side under band bending, we add an extra potential at the surface of the semi-infinite system. The resulting band structure is shown in Fig. 4(d) for $\delta=0.40$. Now we clearly see that a pair of spin-splitting states (magenta dashed lines) cross at the $\bar{\Gamma}$-point. These states are Rashba-like only close to the $\bar{\Gamma}$-point, but evolve close to each other and eventually become spin-degenerate at large momenta, unlike pure Rashba states. The calculated spectral weight of $\delta=0.40$ in Fig. 4(d) is found to be dominated by the outer cone and the spectral weight of the inner cone is strongly suppressed possibly due to scattering with other bulk bands \cite{Helical_metal}, in consistency with our ARPES measurements. The calculated spin polarization of the outer cone follows the left-handed helical configuration as shown in Fig. 4(b), also in consistency with our spin-resolved measurements in Fig. 2.

We have shown the experimental observation and supporting calculation of a spin-momentum locked two dimensional states on the trivial side of the TPT in the BiTl(S$_{1-\delta}$Se$_\delta$)$_2$ system. Calculation implies that such states are spin-splitted trapped bulk states (with inner cone strongly suppressed) cause by surface relaxation due to surface terminated electrical potential. In general, additional trapped bulk states due to band bending are not rarely observed, e.g. on Bi$_2$Se$_3$ surfaces \cite{Andrew Nature physics Fe, Hofmann}. However, two novel properties of the two dimensional states observed here have to be noticed: (i) The measured spectral weight of the two dimensional states arranges in such an exotic way (inner circle completely disappears) that the experimentally measured spin texture of Fermi surface (e.g. $\delta=0.4$ spin data in Fig. 2) resembles the helical spin texture of the topological surface states. (ii) The spin texture on the trivial side is only prominent near the band inversion composition width (e.g. $\delta=0.4-0.5$) but found to be significantly reduced for system far away to the trivial end with large bulk gap ($\delta=0.0$). This strongly indicates that these two dimensional states on the trivial side are correlated with some form of proximity effect from the topological insulator regime across the phase transition. Thus the newly observed spin-momentum locked two dimensional states may serve as a strong precursor which show traces of the topological insulator state even prior to the occurrence of the TPT. In general, these experimental observations provide critical knowledge to form a mature theory of topological phase transition from a trivial insulator to a topological insulator, which is lacking to the date.

Work at Princeton is supported by NSF-DMR-1006492 and the AFOSR MURI on superconductivity. The Synchrotron Radiation Center is supported by NSF-DMR-0537588. M. Z. H. acknowledges Visiting Scientist support from LBNL and additional support from the A. P. Sloan Foundation.

\end{document}